\documentclass[prl,showpacs,amsmath,amssymb,reprint,nofootinbib]{revtex4-1}
\usepackage{graphicx}
\begin{document}
\allowdisplaybreaks[1]
\title{Quantum Gravitational Contributions to the CMB Anisotropy Spectrum}

\author{Claus Kiefer}
\email[E-mail: ]{kiefer@thp.uni-koeln.de}
\affiliation{Institut f\"ur Theoretische Physik, Universit\"at zu
  K\"oln, Z\"ulpicher Stra\ss e 77, 50937 K\"oln, Germany} 

\author{Manuel Kr\"amer}
\email[E-mail: ]{mk@thp.uni-koeln.de}
\affiliation{Institut f\"ur Theoretische Physik, Universit\"at zu
  K\"oln, Z\"ulpicher Stra\ss e 77, 50937 K\"oln, Germany} 

\begin{abstract}
We derive the primordial power spectrum of density fluctuations in the
framework of quantum cosmology. For this purpose we perform a
Born-Oppenheimer approximation to the Wheeler-DeWitt equation for an
inflationary universe with a scalar field. In this way we first
recover the scale-invariant power spectrum that is found as an
approximation in the simplest inflationary models. We then obtain
quantum gravitational corrections to this spectrum and discuss whether
they lead to measurable signatures in the CMB anisotropy spectrum. The
non-observation so far of such corrections translates into an upper
bound on the energy scale of inflation.
\end{abstract}
\maketitle 


Without observational guidance it is illusory to find the correct
quantum theory of gravity. While there exist various approaches to such a
theory (see e.g.~the overview in \cite{oup}), definite predictions
are rare. Among these are the calculation of small quantum gravitational
corrections to the Newtonian and Coulomb potential \cite{BB} and corrections to
Lamb shift and other effects due to the possible existence of a minimal length
\cite{DV08}. While the first effects are too tiny to be observable in
the foreseeable future, the latter depend on a new dimensionless
parameter for which bounds can be found. The calculation of effects is
also important for the comparison of different approaches and for the
decision whether gravity must be quantized at all.  

Our purpose here is to calculate potential observational contributions
to the CMB anisotropy spectrum from quantum gravity. After all, the
main applications of such a theory should arise from
cosmology and black-hole physics. Our framework will be quantum
geometrodynamics governed by the Wheeler-DeWitt equation
\cite{WDW,oup}. Although it is likely that this approach is not the
most fundamental one, one can put forward strong arguments that it 
is approximately valid at energy scales somewhat smaller than the Planck mass
\cite{Kiefer09}. For example, if one looks for a quantum
wave equation that immediately leads to Einstein's
equations in the semiclassical limit, one is directly driven to the
Wheeler-DeWitt equation. 

This connection between the Wheeler-DeWitt equation and quantum field
theory in an external spacetime 
can be established by a Born-Oppenheimer
type of approximation \cite{oup}. Expanding with respect to the Planck
mass, one arrives first
at the functional Schr\"odinger equation for non-gravitational
fields in an external spacetime satisfying the Einstein equations. 
Proceeding with this scheme to the
next orders, one can derive quantum gravitational correction terms
proportional to the inverse Planck-mass squared. The dominating
correction terms are calculated in \cite{KS91} at the formal level of
the full equations.
 The complete set of correction terms at this order
together with their interpretation in terms of Feynman diagrams can be
found in \cite{BK98}.  The generalization of \cite{KS91} to supergravity
is presented in \cite{KLM05}. 

In the present paper we calculate the dominating correction term of
\cite{KS91} for the case of the CMB anisotropy spectrum. In this way
we hope that either a quantum gravitational effect can be observed or
that bounds on cosmological parameters can be found from their
non-observation. 

We shall consider the Wheeler-DeWitt equation for the case of small
fluctuations (leading to the anisotropies in the CMB spectrum) in a flat
Friedmann-Lema\^itre universe with scale factor $a\equiv\exp(\alpha)$
and a scalar field $\phi$ that plays the role of the inflaton.
For definiteness we shall choose
the simplest potential in chaotic inflation \cite{Linde},
$
\mathcal{V}(\phi) = \frac{1}{2}\,m^2\phi^2,
$
but any other potential should fit our purpose as long as at the classical
level a slow-roll condition of the form $\dot{\phi}^2 \ll
\vert\mathcal{V}(\phi)\vert$ holds.
 Setting $\hbar=c=1$, the Wheeler--DeWitt equation for this `minisuperspace
part' reads (see e.g.~\cite{oup})
\begin{eqnarray}
& & {\mathcal H}_0\Psi_0(\alpha,\phi)\equiv \nonumber\\
& & \frac{\text{e}^{-3\alpha}}{2}
\left[\frac{1}{m_{\rm P}^2}\frac{\partial^2}{\partial\alpha^2}
-\frac{\partial^2}{\partial\phi^2}+
\text{e}^{6\alpha}m^2\phi^2\right]\!\Psi_0(\alpha,\phi) 
= 0\,, \label{miniwdw}
\end{eqnarray}
where $m_{\rm P}=\sqrt{3\pi/2G}\approx 2.65 \times 10^{19}$ GeV is a
rescaled Planck mass, 
and the field redefinition $\phi\to\phi/\sqrt{2}\pi$ was performed.

In addition to the Born-Oppenheimer approximation for the
Wheeler-DeWitt equation, we make one further assumption:
we assume that the kinetic term of the $\phi$-field is small compared
to the potential term, that is, $\partial^2\Psi_0/\partial\phi^2\ll 
\text{e}^{6\alpha}m^2\phi^2\Psi_0$.  It corresponds to the slow-roll
approximation for inflationary models and is also the standard
assumption in discussions of the no-boundary and tunneling proposals
in quantum cosmology \cite{oup}; it allows us to neglect the
$\phi$-kinetic term in \eqref{miniwdw}.
For this reason we can also substitute $m\phi$ in
\eqref{miniwdw} by $m_{\rm P}H$, where $H$ is 
the quasistatic Hubble parameter of inflation, which in the classical
limit obeys $\vert\dot{H}\vert\ll H^2$. This replacement of the
quantum variable $\phi$ by a c-number is not problematic here,
because \eqref{miniwdw} describes in the Born-Oppenheimer approximation
the classical background on which the quantum fluctuations of
the inflaton propagate, see below.

We now consider the
fluctuations of an inhomogeneous inflaton field on top of its homogeneous part,
\[
\phi \rightarrow \phi(t) + \delta\phi(\mathbf{x},t)\,,
\]
and perform a decomposition into Fourier modes with wave vector $\mathbf{k}$, $k
\equiv \left|\mathbf{k}\right|$, 
\[
\delta\phi(\mathbf{x},t) =
\sum_k\,f_k(t)\,\text{e}^{\text{i}\mathbf{k}\cdot\mathbf{x}}\,.
\]
(We assume for simplicity that space is compact and the spectrum for 
$\mathbf{k}$ thus discrete.)
The Wheeler-DeWitt equation including the fluctuation modes then
reads \cite{Hall85}
\[
\left[\mathcal{H}_0+\sum_{k=1}^{\infty}
  \mathcal{H}_{k}\right]
  \!\Psi\big(\alpha,\phi,\{f_k\}_{_{k=1}}^{^{\infty}}\big)=0\,, 
\]
where the Hamiltonians $\mathcal{H}_{k}$ of the fluctuation modes are given by
\[
\mathcal{H}_k =
\frac{1}{2}\,\text{e}^{-3\alpha}\left[-\,\frac{\partial^2}{\partial
    f_k^2} + \Bigl(k^2\,\text{e}^{4\alpha} +
  m^2\,\text{e}^{6\alpha}\Bigr)f_k^2\right] .
\]
Since the fluctuations are small, their self-interaction can be
neglected, and one can make the following product ansatz for the full
wave function:
\[
\Psi\big(\alpha,\phi,\{f_k\}_{_{k=1}}^{^{\infty}}\big) =
\Psi_0(\alpha,\phi)\prod_{k=1}^{\infty}\widetilde{\Psi}_k(\alpha,\phi,f_k)\,.
\]
Under some mild assumptions, one finds that the components 
$
\Psi_k(\alpha,\phi,f_k):=
\Psi_0(\alpha,\phi)\widetilde{\Psi}_k(\alpha,\phi,f_k)  
$
obey \cite{Hall85,CK87}
\begin{eqnarray}
\label{wdw}
\frac{1}{2}\,\text{e}^{-3\alpha}\biggl[\frac{1}{m_{\rm
    P}^2}\,\frac{\partial^2}{\partial\alpha^2} &+& 
\text{e}^{6\alpha}\,m_{\rm P}^2\,H^2 \nonumber \\
-\,\frac{\partial^2}{\partial f_k^{2}} &+&
W_k(\alpha)f_k^2\biggr]\Psi_k(\alpha,\phi,f_k) = 0\,, 
\end{eqnarray}
where we have defined the quantity 
\[
W_k(\alpha) := k^2\,\text{e}^{4\alpha} + m^2\,\text{e}^{6\alpha}\, ,
\]
and we have used $m\phi\approx m_{\rm P} H$ as mentioned above.
(For this reason we shall omit the argument $\phi$ in the following.)
Equation \eqref{wdw} 
is the starting point for the Born-Oppenheimer approximation.

Following the general procedure of \cite{KS91}, we make the
ansatz
\[
\Psi_k(\alpha,f_k) = \text{e}^{\text{i}\,S(\alpha,f_k)}
\]
and expand $S(\alpha,f_k)$ in terms of powers of $m_{\rm P}^2$,
\[
S(\alpha,f_k) = m_{\rm P}^2\,S_0 + m_{\rm P}^0\,S_1 + m_{\rm
  P}^{-2}\,S_2 + \ldots . 
\]
Inserting this ansatz into \eqref{wdw} and comparing consecutive
orders of $m_{\rm P}^2$, one obtains at $\mathcal{O}(m_{\rm P}^4)$ that $S_0$ is
independent of $f_k$ and that it obeys at
$\mathcal{O}(m_{\rm P}^2)$ the Hamilton-Jacobi equation
\[
\left[\frac{\partial S_0}{\partial\alpha}\right]^2 
- V(\alpha) =0\,, \quad V(\alpha):={\rm e}^{6\alpha}H^2\, ,
\]
which defines the classical minisuperspace background. Its solution is
$S_0(\alpha)=\pm {\rm e}^{3\alpha}H/3$. 

At $\mathcal{O}(m_{\rm P}^0)$ we first write
\[
\psi^{(0)}_{k}(\alpha,f_{k})\equiv
\gamma(\alpha)\,\text{e}^{\text{i}\,S_{1}(\alpha,f_{k})} 
\]
and impose a condition on $\gamma(\alpha)$ that makes it equal to
the standard WKB prefactor. After introducing the `WKB time' according to
\begin{equation}
\label{WKBtime}
\frac{\partial}{\partial t} :=
-\,\text{e}^{-3\alpha}\frac{\partial S_0}{\partial
    \alpha}\,\frac{\partial}{\partial \alpha}\,,
   \end{equation}
one finds that each $\psi^{(0)}_{k}$ obeys a Schr\"odinger equation,
\begin{equation} \label{Schr_eq}
\text{i}\,\frac{\partial}{\partial t}\,\psi^{(0)}_{k}=
\mathcal{H}_{k} \psi^{(0)}_{k} .
\end{equation}
At the next order
$\mathcal{O}(m_{\rm P}^{-2})$ we decompose $S_2(\alpha,f_k)$ as follows:
\[
S_2(\alpha,f_k)\equiv\varsigma(\alpha)+\eta(\alpha,f_k)
\]
and demand that $\varsigma(\alpha)$ be the standard second-order
WKB correction. 
The wave functions
\[
\psi^{(1)}_{k}(\alpha,f_k) :=
\psi^{(0)}_{k}(\alpha,f_k)\,\text{e}^{\text{i}\,m_{\rm
    P}^{-2}\,\eta(\alpha,f_k)}  
\]
then obey the
quantum gravitationally corrected Schr\"odinger equation \cite{KS91}
\begin{eqnarray} \label{corr_Schr_eq}
& & \text{i}\,\frac{\partial}{\partial t}\,\psi^{(1)}_{k} =
\mathcal{H}_{k}\psi^{(1)}_{k}- \\ & &
\;
\frac{\text{e}^{3\alpha}}{2m_{\rm P}^2\psi^{(0)}_k}\biggl[\frac{\bigl(
\mathcal{H}_{k}\bigr)^2}{V}  
\psi^{(0)}_{k} + \text{i}\frac{\partial}{\partial t}
\left(\frac{\mathcal{H}_{k}}{V}\right)
\psi^{(0)}_{k}\biggr]\psi^{(1)}_k\,. \nonumber 
\end{eqnarray}
In the following we shall only take into account the first correction
term because it usually gives the dominating contribution
\cite{KS91,BK98}. The second correction term corresponds to a small violation of
unitarity, where unitarity is here understood with
respect to the standard ${\mathcal L}^2$-inner product for the modes
$f_k$. While the Hilbert-space structure for full quantum gravity is
unknown \cite{oup}, this is the obvious choice for the $f_k$ because
their states $\psi_k$ obey the approximate Schr\"odinger equation 
\eqref{Schr_eq}. The unitarity-violating term can be absorbed in a
$t$-dependent redefinition of the states \cite{CL95}. 

We shall now first look for a solution of the
uncorrected Schr\"odinger equation (\ref{Schr_eq}). We make a
Gaussian ansatz,
\begin{equation}
\label{Gaussianansatz}
\psi^{(0)}_{k}(t,f_k) =
\mathcal{N}_k^{(0)}(t)\,\text{e}^{-\frac{1}{2}\,\Omega_k^{(0)}(t)\,f_k^2} .
\end{equation}
Here, we have expressed $\alpha$ in terms of the WKB time $t$
introduced in \eqref{WKBtime}, $\alpha=Ht$. We thereby arrive at the following
system of differential equations:
\begin{eqnarray}
\dot{\mathcal{N}}_k^{(0)}(t) &=&
-\,\frac{\text{i}}{2}\,\text{e}^{-3\alpha}\,\mathcal{N}_k^{(0)}(t)\,
\Omega_k^{(0)}(t) , \label{prefactor}
\\ 
\dot{\Omega}_k^{(0)}(t) &=&
\text{i}\,\text{e}^{-3\alpha}
\Bigl[-\bigl(\Omega_k^{(0)}(t)\bigr)^2+W_k(t)\Bigr]. \label{Omegazero}
\end{eqnarray}
In the model of chaotic inflation employed here we have the condition
$(m/H)^2\ll 1$ \cite{Linde}. In this limit the solution 
of \eqref{Omegazero} expressed in
terms of the dimensionless quantity $\xi(t) := k/(Ha(t))$ reads  
\begin{equation} \label{Sol_Omega0}
\Omega^{(0)}_k(\xi) = \frac{k^3}{H^2\xi}\,\frac{1}{\xi-\text{i}} +
\mathcal{O}\!\left(\frac{m^2}{H^2}\right) . 
\end{equation} 
 From \eqref{prefactor} and the normalization of the states
one then obtains the solution $\vert\mathcal{N}_k^{(0)}(t)\vert^2
=(\Re{\frak e}\,\Omega_k^{(0)}(t)/\pi)^{1/2}$.

In the slow-roll regime, the density contrast is given by (see
e.g.~\cite{Paddy}, p.~364) 
\[
\delta_k(t) \approx \frac{\delta{\rho}_k(t)}{\mathcal{V}_0} =
\frac{\dot{\phi}(t)\,\dot{\sigma}_k(t)}{\mathcal{V}_0}\,,
\]
where $\mathcal{V}_0$ denotes the scalar-field potential evaluated at
the background solution $\phi(t)$,
and $\sigma_k(t)$ is the classical quantity related to the quantum
mechanical variable $f_k(t)$ by taking its expectation value with
respect to a Gaussian state; for a general Gaussian we define
\begin{eqnarray*}
& &\sigma_k^2(t) := \left\langle\psi_k\vert f_k^2\vert\psi_k
\right\rangle \\
&=&\sqrt{\frac{\Re{\frak e}\, \Omega_k}{\pi}}\int\limits_{-\infty}^{\infty}
f_k^2\, \text{e}^{-\frac{1}{2}\left[\Omega^*_k(t)+\Omega_k(t)\right]f_k^2}
\,\text{d} f_k  
= \frac{1}{2\,\Re{\frak e}\,\Omega_k(t)}\,.
\end{eqnarray*}
The density contrast must be
evaluated at the time $t_{\text{enter}}$ when the corresponding mode
re-enters the Hubble radius during the radiation-dominated phase. A
standard relation gives (\cite{Paddy}, p.~367) 
\[
\delta_k(t_{\text{enter}}) =
\frac{4}{3}\,\frac{\mathcal{V}_0}{\dot{\phi}^2}\,\delta_k(t_{\text{exit}})
=
\frac{4}{3}\,\frac{\dot{\sigma}_k(t)}{\dot{\phi}(t)}\Biggl|_{t\,=\,t_{\text{exit}}} 
.
\]
Evaluating $\dot{\sigma}_k^{(0)}(t)$ at $t = t_{\text{exit}}$ using
(\ref{Sol_Omega0}) and noting that
$
\xi(t_\text{exit}) = 2\pi
$
at Hubble-scale crossing, we get 
\[
\left|\dot{\sigma}_k^{(0)}(t)\right|_{t\,=\,t_{\text{exit}}} =
 \frac{2\sqrt{2}\,\pi^2}{\sqrt{4\pi^2 +1}}\,\frac{H^2}{k^{\frac{3}{2}}}\,.
\]
This then leads to the power spectrum
\begin{equation}
\label{power}
\Delta^{\!2}_{(0)}(k) := 4\pi
k^3\left|\delta_k(t_{\text{enter}})\right|^2 \propto
\frac{H^4}{\big|\dot{\phi}(t)\big|^2}_{\!\!\!\!t_{\text{exit}}} ,
\end{equation} 
which is approximately scale-invariant. 
This is the standard result for a generic inflationary model.

We now want to calculate the quantum gravitational correction terms 
following from (\ref{corr_Schr_eq}). (A possible
  effect on the relic graviton density is discussed along these
  lines in \cite{Rosales97}.) 
As mentioned above, we shall neglect the
unitarity violating term in \eqref{corr_Schr_eq}. We assume that the
correction can be accommodated by the Gaussian ansatz
\begin{eqnarray*}
\ \ \psi^{(1)}_{k}(t,f_k) &=& \left(\mathcal{N}_k^{(0)}(t) +
  \frac{1}{m_{\rm P}^2}\,\mathcal{N}_k^{(1)}(t)\right) \\ 
&\times&
\exp\!\left[-\,\frac{1}{2}\left(\Omega_k^{(0)}(t)+\frac{1}{m_{\rm
        P}^2}\,\Omega_k^{(1)}(t)\right)f_k^2\right]  .
\end{eqnarray*}
 One then gets from \eqref{corr_Schr_eq}
an equation for the correction term $\Omega_k^{(1)}$,
\begin{eqnarray}
\label{Omegacorr}
\dot{\Omega}_k^{(1)}(t) \approx
&-& 2\,\text{i}\,\text{e}^{-3\alpha}\,\Omega_k^{(0)}(t)\times \\
&\;&\left(\Omega_k^{(1)}(t)
- \frac{3}{4V(t)}\left[\bigl(\Omega_k^{(0)}(t)\bigr)^2 -
  W_k(t)\right]\right)\,.\nonumber 
\end{eqnarray}
We shall assume that the correction term vanishes for late times, 
$
\Omega_k^{(1)}(t) \rightarrow 0 \;\text{ as }\; t \rightarrow \infty.
$
This is, of course, an assumption that must eventually be justified
from the theory itself;
the chosen boundary condition guarantees
that the model is consistent and in accordance with observations at
late times.

Using \eqref{Sol_Omega0}, we rewrite \eqref{Omegacorr} in terms of
$\xi$, which in the limit $(m/H)^2 \ll 1$ gives
\begin{equation}
\label{OmegacorrXi}
\frac{\text{d}}{\text{d}\xi}\,\Omega_k^{(1)}(\xi) =
\frac{2\,\text{i}\,\xi}{\xi-\text{i}}\,\Omega_k^{(1)}(\xi) +
\frac{3\,\xi^3}{2}\,\frac{2\xi-\text{i}}{(\xi-\text{i})^3}\,. 
\end{equation}
The corrected quantity $\dot{\sigma}_k^{(1)}$ needed for the
evaluation of the power spectrum \eqref{power} is then given by
\begin{eqnarray*}
& & \bigl|\dot{\sigma}_k^{(1)}(t)\bigr| =
\Biggl|\frac{H\xi}{\sqrt{2}}\,\frac{\text{d}}{\text{d}
  \xi}\!\left[\!\left(\Re\mathfrak{e}\,\Omega^{(0)}_k(\xi) +
    \frac{1}{m_{\rm P}^2}\,\Re{\frak 
      e}\,\Omega_k^{(1)}(\xi)\right)^{\!-\frac{1}{2}}\right]\!\Bigg|
\\ 
\ \ & & =
\Biggl|\frac{\xi^2}{\sqrt{2(\xi^2+1)}}\,\frac{H^2}{k^{\frac{3}{2}}}\left(1
  +
  \frac{\xi^2+1}{k^3}\,\Re\mathfrak{e}\,\Omega_k^{(1)}(\xi)\,\frac{H^2}{m_{\rm 
      P}^2}\right)^{\!-\frac{3}{2}}     
 \\  
& & \ \ \ \times\left(1 -
  \frac{(\xi^2+1)^2}{2\xi\,k^3}\,\Re\mathfrak{e}\!\left[\frac{\mathrm{d}}{\mathrm{d}  
      \xi}\,\Omega_k^{(1)}(\xi)\right]\frac{H^2}{m_{\rm P}^2}\right)\!\Bigg|\,.
\end{eqnarray*}
The solution of \eqref{OmegacorrXi} can be reduced to 
numerical integration and yields $\Re\mathfrak{e}\,\Omega_k^{(1)}(\xi =
2\pi) \simeq -1.076$ as well as
$\Re\mathfrak{e}\!\left[\mathrm{d}\Omega_k^{(1)}(\xi)/\mathrm{d}\xi\right]_{\xi
  = 2\pi} \simeq 1.451$, which eventually leads to
\begin{equation}
\label{sigmacorr}
\bigl|\dot{\sigma}_k^{(1)}\bigr|_{t_{\text{exit}}}\simeq \vert C_k\vert
\bigl|\dot{\sigma}_k^{(0)}\bigr|_{t_{\text{exit}}} ,
\end{equation}
where
\begin{equation}
\label{Ck}
C_k:=\left(1 -
    \frac{43.56}{k^3}\,\frac{H^2}{m_{\text{P}}^2}\right)^{\!\!-\frac{3}{2}}
  \!\!\left(1 
  - \frac{189.18}{k^3}\,\frac{H^2}{m_{\text{P}}^2}\right).
\end{equation}
With this result we can write the corrected power spectrum as the
product of the  
uncorrected power spectrum with a correction term $C_k$,
$
\Delta^{\!2}_{(1)}(k) = \Delta^{\!2}_{(0)}(k)\,C^2_k\,.
$
An expansion of $C_k^2$ in terms of $(H/m_{\rm P})^2$ yields
\begin{eqnarray}
\label{Delta1}
\Delta^{\!2}_{(1)}(k) &\simeq&
\Delta^{\!2}_{(0)}(k) \nonumber \\ &\times& 
\left[1-\frac{123.83}{k^3}\,\frac{H^2}{m_{\text{P}}^2}+
  \frac{1}{k^6}\,{\mathcal
    O}\!\left(\frac{H^4}{m_{\text{P}}^4}\right)\right]^2. 
\end{eqnarray}
We emphasize the important fact that the corrected power spectrum is
now explicitly scale-dependent. The quantitative contribution of the
quantum gravitational terms is only significant if the inflationary
Hubble para\-me\-ter $H$ is sufficiently large. It is not surprising that
the effects become sizeable only if $H$ approaches the Planck scale. 

An inspection of \eqref{Ck} shows that $C_k$ approaches one for large
$k$ (as it must), but decreases monotonically to zero for large scales (small
$k$); 
one thus finds a {\em suppression of power} for large scales. The zero
point is reached for $k\approx 5.74 (H/m_{\rm P})^{2/3}$. However, the 
approximation \eqref{Ck} breaks down if this zero point is approached
and one has to take into account in this limit
higher orders of $(H/m_{\rm P})^2$.

The effect is most prominent for large scales because these scales are
the earliest to leave the Hubble scale during inflation. However, the
measurement accuracy for large scales is fundamentally limited by
cosmic variance, which follows from the fact that we only observe one
Universe (see e.g.~\cite{PU}). For this reason, missions such as the 
PLANCK satellite will not be able to see this effect if it has not
already been seen now. But there is still a merit of our analysis:
from the current
non-observation of the quantum gravity terms one can get an upper
bound on the inflationary Hubble scale. Assuming for a rough estimate
that $C_k^2$ is not less 
than around 0.95 for the largest observable scales $k\sim 1$
(which is motivated by the fact that the deviation of
the observed power spectrum from a scale-invariant spectrum is smaller
than about $5\ \%$ \cite{Koma11}), one obtains from \eqref{Delta1} the bound
\begin{equation}
\label{Hbound}
H \lesssim 1.4\times 10^{-2}\,m_{\text{P}} \sim
4\times10^{17}\,\text{GeV}\,.
\end{equation}
We must emphasize, however, that there already exists a stronger
constraint on this scale. This is because
the energy scale of inflation is limited by the observational
bound on the tensor-to-scalar ratio $r$ (see
e.g.~\cite{CMBPol}). Using $r<0.22$ \cite{Koma11} one finds $H\lesssim
10^{-5}m_{\text{P}}\sim 10^{14}\,\text{GeV}$. As emphasized, for
example, in \cite{alexei}, the assumption $H\ll m_{\rm P}$ is anyway
required and self-consistent for inflationary models to have a
connection with reality. 
For the limiting value $H\sim 10^{14}\,\text{GeV}$ one gets from
\eqref{Delta1}   
\begin{equation*}
\Delta^{\!2}_{(1)}(k) \simeq
\Delta^{\!2}_{(0)}(k)\left[1-1.76\times 10^{-9}\,\frac{1}{k^3}
+ \frac{{\mathcal O}(10^{-15})}{k^6}\right]^2 ;
\end{equation*}
for this value the correction is thus too small to be seen in present
observations. 

In spite of this, we emphasize that our constraint \eqref{Hbound}
arises as a definite prediction from a conservative approach to quantum
gravity, and it is reassuring that it is consistent with other
limits. It indicates, in particular, that no additional trans-Planckian effects
(see e.g.~\cite{alexei,PU}) have to be taken into account in order to
understand the predictions of this model.

Quantum gravitational corrections to the CMB anisotropy spectrum have
also been derived in loop quantum cosmology. While in
 \cite{Tsuj04} a suppression of power at large scales was found,
the authors of \cite{BCT} predicted an enhancement at those scales.
This demonstrates that one can use the CMB anisotropies to compare 
different approaches to quantum 
gravity. We hope that such investigations will eventually
lead to an observational test of quantum gravity.

\vskip 2mm
We are grateful to Christian Steinwachs for interesting discussions
and critical comments.
M.~K. acknowledges support from the Bonn-Cologne Graduate School of Physics and
Astronomy. He also thanks Giulio Rampa for valuable discussions and remarks.


\end{document}